\documentclass[twocolumn,tighten]{aastex62}
\usepackage{amsmath}

\usepackage{xcolor}
\usepackage{verbatim}

\shorttitle{Disruption of Spinning Stars}
\shortauthors{Golightly et al.}

\begin{document}

\title{Tidal disruption events: the role of stellar spin}

\correspondingauthor{Elen Golightly}
\email{ecag2@le.ac.uk}

\author{Elen~C.~A.~Golightly}
\affiliation{Department of Physics and Astronomy, University of Leicester, Leicester, LE1 7RH, UK}

\author[0000-0003-3765-6401]{Eric~R.~Coughlin}
\altaffiliation{Einstein Fellow}
\affiliation{Columbia Astrophysics Laboratory, Columbia University, New York, NY, 10027, USA}

\author[0000-0002-2137-4146]{C.~J.~Nixon}
\affiliation{Department of Physics and Astronomy, University of Leicester, Leicester, LE1 7RH, UK}

\begin{abstract}
The tidal force from a supermassive black hole can rip apart a star that passes close enough in what is known as a Tidal Disruption Event. Typically half of the destroyed star remains bound to the black hole and falls back on highly eccentric orbits, forming an accretion flow which powers a luminous flare. In this paper we use analytical and numerical calculations to explore the effect of stellar rotation on the fallback rate of material. We find that slowly spinning stars ($\Omega_* \lesssim 0.01 \Omega_{\rm{breakup}}$) provide only a small perturbation to fallback rates found in the non-spinning case. However when the star spins faster, there can be significant effects. If the star is spinning retrograde with respect to its orbit the tidal force from the black hole has to spin down the star first before disrupting it, causing delayed and sometimes only partial disruption events. However, if the star is spinning prograde this works with the tidal force and the material falls back sooner and with a higher peak rate. We examine the power-law index of the fallback curves, finding that in all cases the fallback rate overshoots the canonical $t^{-5/3}$ rate briefly after the peak, with the depth of the overshoot dependent on the stellar spin. We also find that in general the late time evolution is slightly flatter than the canonical $t^{-5/3}$ rate. We therefore conclude that considering the spin of the star may be important in modelling observed TDE lightcurves.
\end{abstract}


%%%%%%%%%%%%%%%%%%%%%%%%%%%%%%%%%%%%%%%%%%%%%%%%%%

%%%%%%%%%%%%%%%%% BODY OF PAPER %%%%%%%%%%%%%%%%%%

\section{Introduction}

If a star passes close enough to a supermassive black hole in the centre of a galaxy, it can be stretched and pulled apart by the hole's gravitational field in what is known as a tidal disruption event (TDE). To be disrupted the star must pass within the tidal radius, $r_{\rm{t}}=\big(M_{\rm{BH}}/{M_*} \big)^{1/3}R_{*}$, which is the distance at which the tidal force from the black hole overcomes the self-gravity of the star. Some of the stellar debris is expelled into the galaxy, while the rest falls back towards the black hole and is expected to circularize into an accretion disc, losing energy and feeding the hole to generate a high energy flare \citep{rees_tidal_1988}. The appearance of the flares depends on several parameters including black hole mass and spin, stellar properties and the orbit of the star. The amount of mass falling back to the accretion disc over time can be inferred from the lightcurves which show a rise to a peak and then a power law decay. This decay typically takes the form $t^{-\frac{5}{3}}$ \citep{rees_tidal_1988, phinney_manifestations_1989}, derived from the negative Kepler energy of the bound debris:

\begin{equation}\label{eq:Ebound}
	E_{\rm{bound}}=-\frac{1}{2} \bigg(\frac{2\pi G M_{\rm{BH}}}{t}\bigg)^\frac{2}{3}
\end{equation}
and with the specific energy distribution $\frac{dM}{dE}$ assumed to be uniform. This shows the rate of fallback is proportional to $t^{-\frac{5}{3}}$:

\begin{equation}
	\dot{M}_{\rm{fb}}=\frac{dM}{dt}=\frac{dM}{dE}\frac{dE}{dt}=\frac{dM}{dE}\frac{1}{3}(2\pi G M_{\rm{BH}})^{\frac{2}{3}}t^{-\frac{5}{3}}
\end{equation}

TDEs are usually modelled with the star on a parabolic orbit as most tidally disrupted stars are assumed to come from large radii and need to reach pericentre at $\sim 50R_{\rm{g}}$, where $R_{\rm{g}}$ is the gravitational radius $R_{\rm{g}}=GM_{\rm{BH}}/c^2$, giving the eccentricity of the orbit as close to unity: $e=\frac{r_{\rm{a}}-r_{\rm{p}}}{r_{\rm{a}}+r_{\rm{p}}}\approx 1$. As a result of the orbit being parabolic the amount of bound/unbound material is distributed $50-50$.

The star reaches the tidal radius with an impact parameter, $\beta$, the ratio of tidal radius to pericentre distance. The extent of the disruption event depends on this impact parameter as follows: the critical point of disruption is at $\beta_{\rm{c}} = 1$, where a larger $\beta$ gives a deeper orbit, and the star is only partially disrupted if $\beta < 1$, and no disruption occurs for $\beta \ll 1$. However, $\beta_{\rm{c}}$ is dependent on stellar structure and can vary by a factor of $\sim2$ \citep{guillochon_hydrodynamical_2013}. $\beta_{\rm{c}}$ would also vary with the spin of the star depending on spin magnitude and direction.

A star spins at a fraction of its break up velocity which in addition to the black hole's tidal force would quicken its disruption, provided they were in the same direction. If the star is spinning against the direction of the tidal force, it would hinder the disruption and can leave behind an intact portion of the star.

\cite{lodato_stellar_2009} showed that the lightcurve is only proportional to $t^{-\frac{5}{3}}$ at late times and depends on stellar structure during the early stage of the fallback. They show that more compressible stars (smaller values of $\gamma$, the polytropic index) give a more gentle rise to the peak. They conclude that the $t^{-\frac{5}{3}}$ decay only holds where the energy distribution $dM/dE$ is constant which is approximately true only at late times. Given that most of the accreted material returns to the black hole before reaching $t^{-\frac{5}{3}}$, it is not clear how many observed events would be characterised by a $t^{-\frac{5}{3}}$ lightcurve. At late times the lightcurve is expected to change shape due to the viscous timescale in the accretion disc \citep[power law index $\sim-1.2$;][]{cannizzo_disk_1990}.

\cite{guillochon_hydrodynamical_2013} looked at the effect of changing the impact parameter for cases spanning grazing encounters (low $\beta$) to deep plunges (high $\beta$) using $\gamma=4/3,5/3$ for high and low mass main sequence stars. They showed that the most concentrated stars drop in fallback rate quickly after the peak where only partial disruption of the star occurred. Their results show much steeper decays than the expected $t^{-\frac{5}{3}}$ where some of the stellar core remains intact, which suggests we can determine from observations if a flare has produced a survived core.

A star will break up if it is spun up to the point at which $F_{\rm{sg}} \sim F_{\rm{rot}}$ where self-gravity of the star is ${GM^2_*}/{R^2_*}$ and the centrifugal force is $\pm M_*\Omega^2_{\rm{*}}R_*$, where $\Omega_{\rm{*}}$ is the stellar rotation frequency. This gives a break up angular velocity:
\begin{equation}
\Omega_{\rm{br}}= \sqrt{\frac{GM_*}{R^3_*}}
\end{equation}
We define the break-up fraction as $\lambda=\Omega_{\rm{*}} /\Omega_{\rm{br}}$ where $\Omega_{\rm{*}}>0$ is a prograde spin and $\Omega_{\rm{*}}<0$ is retrograde.

Current observations of spinning stars include the Sun which spins at $0.002$ of its break up velocity (with an equatorial period of $24$ days) and \cite{meibom_spin-down_2015} who found in a sample of $30$ cool stars, from the NGC 6819 cluster observed by Kepler, they spun on average with $0.035\Omega_{\rm{br}}$ (assuming a stellar radius, from a rotation period of $18.2$ days) and \cite{nascimento_jr_rotation_2014} found a mean rotation period of $19$ days over $43$ main-sequence stars but with some as slow as $27$ days. \cite{mcquillan_rotation_2014} determined the periods for $34,030$ Kepler MS stars of temperature $<6500$K and found a range of periods from $0.2$ to $70$ days. Assuming Sun-like mass and radius, this faster period corresponds to a spin $0.057 \Omega_{\rm{br}}$. However, given the bias of TDE detections in post-starburst galaxies and, as young stars rotate faster, we might expect many disruptions involving stars with higher spins. For example, a solar mass Pre-Main Sequence star of age $\sim 1Myr$ would reach spins of $0.07$ of break up \citep{bouvier_observational_2013}. A B-type MS star, HD43317, observed by CoRoT rotates at $50\%$ of its critical velocity which corresponds to $\sim 0.28 \Omega_{\rm{br}}$ \citep{rieutord_two-dimensional_2013}.

TDEs are produced and fill the loss cone via one of two regimes: the diffusive or pinhole. In the pinhole regime a far away star experiences one large kick to send it on an orbital path passing within the tidal radius of a SMBH. These stars pass within the loss cone multiple time in an orbit \citep{stone_rates_2016}. For stars much closer, the loss cone is filled slowly by stars diffusing over many stellar orbits. This diffusive regime can cause  stars to be slowly pushed to smaller orbits and then the tidal field can spin up the star to larger prograde velocities before disruption occurs. Stars in the pinhole regime tend to have large impact parameters whereas diffusive regime stars have $\beta \approx 1$ \citep{stone_rates_2016}.

With the first light of LSST due in 2021, which it will do all-sky surveys every 3 days \citep{lsst_science_collaboration_science-driven_2017}, the catalogue of observed TDEs is expected to rapidly increase. Therefore more accurate lightcurves would be useful to classify new observations. In this work we look at the whether stellar spin has a significant role in numerical TDE models. We look at both prograde and retrograde spins and compare their fallback rate curves to the non-spinning case, as well as analysing how the decay of the curves compare with predictions. Previous models of tidal disruption events have neglected the use of spinning stars so it is important to test whether it is actually fair to exclude this. Fallback curves are so far only constrained to a few parameters that determine the stream evolution, such as impact parameter, mass ratio and polytropic index. With the inclusion of other properties, such as stellar spin, we may find that there is a larger error involved when constraining these parameters. 

\cite{stone_consequences_2013} show that stellar spin widens the energy distribution of the star and note that the misalignment between large stellar spins and the orbital angular momentum could have a larger effect on the energy spread. We can estimate the tidal radius for a spinning star \citep[cf.][]{kesden_tidal_2012} by balancing the self-gravity with the tidal force and force due to spin:

\begin{equation}
F_{\rm{sg}}=F_{\rm{tidal}}\pm F_{\rm{rot}}
\end{equation}

\begin{equation}
\frac{GM^2_*}{R^2_*}=\frac{GM_{\rm{BH}}M_*R_*}{r^3_{\rm{t}}}\pm M_*\Omega^2_{\rm{*}}R_*
\end{equation} The $\pm$ allows for prograde and retrograde spins where prograde ($+$) helps the tidal force and retrograde ($-$) hinders.

\begin{equation}\label{eq:rtspin}
r_{\rm{t}}(\Omega)=R_* \bigg( \frac{M_{\rm{BH}}}{M_*} \bigg)^{\frac{1}{3}} \bigg( 1 \mp \frac{\Omega^2_{\rm{*}}R^3_*}{GM_*}\bigg)^{-\frac{1}{3}}
\end{equation}

We note that this estimate behaves appropriately in the correct limits. When $\Omega_{\rm{*}}=0$ it recovers the standard tidal radius and when $\Omega_{\rm{*}}=+\Omega_{\rm{br}}$ the radius is infinite, and when $-\Omega_{\rm{br}}$ the radius is reduced appropriately. 

In Section 2 we present analytical calculations. In Section 3 we present our numerical simulations, and present our conclusions in Section 4.

\section{Analytical Predictions} \label{analyticalprediction}
\subsection{Impulse approximation}
Assume that the tidal force acts impulsively (cf.~\citealt{lodato_stellar_2009}), meaning that the star is unperturbed and maintains perfect hydrostatic balance prior to reaching the tidal radius, and is completely destroyed (self-gravity and pressure negligible) after the star passes through the tidal radius. Under this ``impulse approximation,'' the energy of a given gas parcel within the star at the moment the stellar center of mass (COM) reaches the tidal radius is

\begin{equation}
\epsilon = \frac{1}{2}\mathbf{v}^2-\frac{GM}{|\mathbf{r}|},
\end{equation}
where $\mathbf{v}$ is the instantaneous velocity vector of the gas parcel and $\mathbf{r}$ is its vector displacement from the black hole. Since the star is assumed to be in hydrostatic equilibrium, we can write

\begin{equation}
\mathbf{v} = \mathbf{v}_*+\mathbf{\Omega}_*\times \mathbf{R},
\end{equation}
where $\mathbf{v}_*$ is the stellar COM velocity, $\mathbf{\Omega}_*$ is the angular velocity of the star, and $\mathbf{R}$ is the position of the gas parcel within the star measured from the stellar COM. Further writing $\mathbf{r} = \mathbf{r}_*+\mathbf{R}$, where $\mathbf{r}_*$ is the position of the COM, employing the tidal approximation (i.e., keeping only first-order terms in the small quantity $R/r_*$), and using the fact that the star is on a parabolic orbit, the energy becomes

\begin{equation}
\epsilon = \left(\mathbf{v}_*\times\mathbf{\Omega}_*+\frac{GM}{r_*^3}\mathbf{r}_*\right)\cdot\mathbf{R}+\frac{1}{2}\left(\mathbf{\Omega}_*\times \mathbf{R}\right)^2 \label{Eeq}
\end{equation}
This expression yields the (conserved) energy of a gas parcel as a function of its position within the star, $\mathbf{R}$, following the disruption, and from it we can deduce a number of important effects of rotation on the dynamics of the TDE.

For one, the lowest-order (in $\mathbf{\Omega_*}$) effect of rotation is given by the first term in parentheses in this equation, which scales identically with $\mathbf{R}$ as the standard (non-spinning) energy spread induced from the tidal force of the black hole, being the second term in parentheses. Equating these two yields the rotation rate of the star that would match the spread in energy generated by the tidal force alone. Defining this rotation rate as $\Omega_{\rm eq} = \lambda_{\rm eq} \Omega_{\rm br}$, where $\Omega_{\rm br} = \sqrt{GM_*/R_*^3}$ is the breakup rotation rate of the star ($M_*$ and $R_*$ are the stellar mass and radius, respectively), then performing some algebra demonstrates that $\lambda_{\rm eq} \simeq 1$. Therefore, if the star is rotating at only a small fraction of its breakup speed, then the spread in energy imparted by rotation will always be subdominant to that imparted by the tidal force. The nonlinear term in Equation \eqref{Eeq} is, therefore, negligible for all physical stars.

Second, if the angular velocity is exactly parallel to the center of mass velocity, then there is no first order term in the correction to the energy spread and the effects of rotation on the evolution of the tidally-disrupted debris are much smaller. In this case, the centrifugal barrier generated by the rotation would reduce the gravitational self-confinement of the tidally-disrupted debris stream. The ability of the stream to fragment under its own self-gravity, as seen in \citet{coughlin_variability_2015} and \citet{coughlin_post-periapsis_2016}, would then be inhibited, but the spread in the energy along the stream would be largely unaltered by the rotation.

Third, if the star is spinning at an oblique angle relative to the COM velocity, then Equation \eqref{Eeq} becomes (assuming that the rotation rate is sub-breakup and ignoring the nonlinear term):

\begin{equation}
\epsilon = \left(\left[v_*\Omega_{\rm z}+\frac{GM}{r_*^2}\right]\sin\theta\cos\phi-v_*\Omega_{\rm x}\cos\theta\right)R. \label{Eeq2}
\end{equation}
In this expression, we let the line connecting the black hole and the stellar COM define the $x$-direction, the stellar COM velocity is in the $y$-direction, and the $z$-direction is defined from these in a right-handed sense; the rotation rate of the star is then $\mathbf{\Omega}_* = \left\{\Omega_{\rm x},\Omega_{\rm y},\Omega_{\rm z}\right\}$ in these coordinates. We also defined the position of the gas parcel within the star in spherical coordinates, so $\mathbf{R} = \left\{R\sin\theta\cos\phi,R\sin\theta\sin\phi,R\cos\theta\right\}$. 

Interestingly, Equation \eqref{Eeq2} demonstrates that rotation in the $x$-direction serves to tilt the location of the most-bound debris element out of the orbital plane of the star: differentiating Equation \eqref{Eeq2} with respect to $\theta$, where $\phi = \pi$ for the most bound material, the energy is minimized at the angle

\begin{equation}
\cot\theta_{\rm m} = \frac{v_*\Omega_{\rm x}}{v_*\Omega_{\rm z}+\frac{GM}{r_*^2}}. \label{thm}
\end{equation}
Since $v_*\Omega_{\rm z} \ll GM/r_*^2$ when the star is rotating at sub-breakup velocities, we can Taylor expand this equation to give

\begin{equation}
\theta_{\rm m} \simeq \frac{\pi}{2}-\sqrt{2}\lambda_{\rm x},
\end{equation}
where $\lambda_{\rm x}$ is defined by $\Omega_{\rm x} = \lambda_{\rm x}\sqrt{GM_*/R_*^3}$. It is also straightforward to show that the most-unbound debris element is located at an angle of $\pi/2+\sqrt{2}\lambda_{\rm x}$. Therefore, rotation in the $x$-direction shifts the line defining the maximum energy spread from the $x$-axis to one that is tilted from the $x$-axis by the small angle $\sqrt{2}\lambda_{\rm x}$. 

Finally, if the angular velocity is purely in the $z$-direction, then $\mathbf{v}_*\times\mathbf{\Omega} \propto \mathbf{r}_*$, and the expression for the energy simplifies to

\begin{equation}
\epsilon = \left(v_*\Omega+\frac{GM}{r_*^2}\right)x = \left(1+\sqrt{2}\lambda\right)\frac{GMx}{r_*^2}, \label{Epar}
\end{equation}
where $x = R_*\sin\theta\cos\phi$ and the last line follows from letting $v_* = \sqrt{2GM/r_*}$, $r_* = R_*(M/M_*)^{1/3}$, and $\Omega = \lambda\sqrt{GM_*/R_*^3}$. This expression demonstrates that, as for the non-rotating case, surfaces of constant energy within the star coincide with surfaces of constant linear displacement in the direction connecting the black hole and the stellar COM. We also see that, if $\Omega$ is positive, corresponding to alignment between the orbital angular momentum vector of the stellar COM and the angular velocity of the star, the total spread in the energy increases, while antialignment reduces the effective energy spread. Thus, stars rotating in a prograde sense -- $\Omega$ aligned with the angular momentum of the star -- are more easily disrupted than non-spinning stars, while retrograde-rotating stars are less easily disrupted. 

Equation \eqref{Epar} also shows that the most-bound gas parcel has an energy

\begin{equation}
\epsilon_{\rm mb} = -\left(1+\sqrt{2}\lambda\right)\frac{GMR_*}{r_*^2}. \label{Emb}
\end{equation}
Using the energy-period relation for a Keplerian orbit then gives the return time of the most tightly bound debris:

\begin{equation}
T_{\rm mb} = \left(\frac{R_*}{2}\right)^{3/2}\frac{2\pi M}{M_*\sqrt{GM}}\left(1+\sqrt{2}\lambda\right)^{-3/2}.
\end{equation}
This shows that prograde-spinning stars (positive $\lambda$) result in shorter return times of the most-bound debris, while retrograde spin yields a longer return time. 

\subsection{Fallback model}
Here we construct a simple, analytic model for the return of the debris to the black hole following the disruption of a spinning star analogous to that presented in \citet{lodato_stellar_2009}. For the sake of simplicity and because our numerical simulations are restricted to this case (see Section \ref{sec:simulations}), we confine our calculations to the scenario in which the stellar angular velocity is parallel to the orbital angular momentum of the star, so Equation \eqref{Epar} describes the spread in the energy following the disruption. Tilts to the rotation should only slightly modify these results, as shown in the previous subsection.

From Equation \eqref{Epar}, the energy of a given gas parcel within the star at the time of disruption is purely a function of its linear position in the star, $x$. Furthermore, the energy-period relation of a Keplerian orbit ensures that all gas parcels with the same energy return to the black hole at the same time $t$. Therefore, to derive the total fallback mass $M_{\rm fb}$ that has yet to return to the black hole, one can parameterize the mass in the star at the time of disruption in terms of $x$ and then use Equation \eqref{Epar} to write $x(\epsilon) = x(\epsilon(t))$. It follows geometrically that this can be written (see also \citealt{lodato_stellar_2009} and \citealt{coughlin14})

\begin{equation}
dM_{\rm fb}(x) = 2\pi \int_x^{R_*}\rho(R)RdRdx, \label{dMfb}
\end{equation}
where $\rho$ is the density of the star at the time of disruption, assumed to be the original stellar density profile; by writing $\rho(R)$ we are ignoring any latitudinal variations induced by the stellar rotation. This is a good approximation when the rotation rate is not too close to the breakup velocity. Using the energy-period relation for a Keplerian orbit in Equation \eqref{dMfb} then gives for the fallback rate

\begin{equation}
\frac{dM_{\rm fb}}{dt} \equiv \dot{M}_{\rm fb} = \frac{4\pi}{3}\frac{R_*}{T_{\rm mb}}\left(\frac{t}{T_{\rm mb}}\right)^{-5/3}\int_{x(t)}^{R_*}\rho(R)RdR,
\end{equation}
where $x(t) = R_*\left(t/T_{\rm mb}\right)^{-2/3}$. Finally, if we define the dimensionless position within the star as $\eta = R/R_*$ and the average stellar density by $\rho_* = 3M_*/(4\pi R_*^3)$, then this expression simplifies to

\begin{equation}
\dot{M}_{\rm fb} = \frac{M_*}{T_{\rm mb}}\left(\frac{t}{T_{\rm mb}}\right)^{-5/3}\int_{x(t)/R_*}^{1}\frac{\rho\eta d\eta}{\rho_*}. \label{mdotfb}
\end{equation}
This expression clearly illustrates the rough magnitude and asymptotic, $t^{-5/3}$ scaling of the fallback rate.

\begin{figure*}
	\centering
	\includegraphics[width=0.7\textwidth]{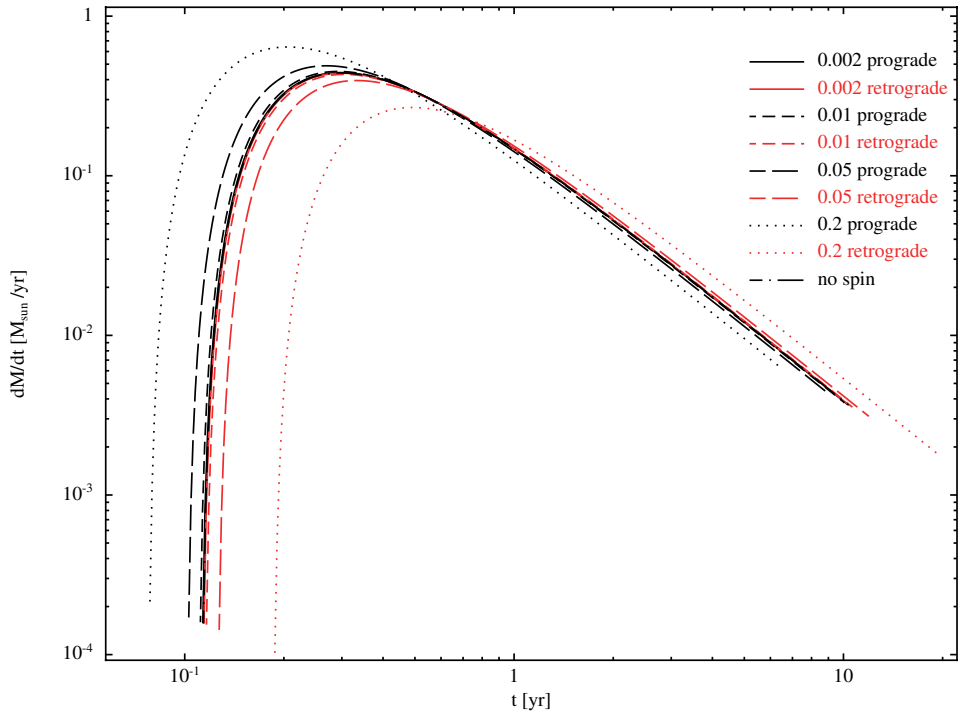}
    \caption{{The predicted fallback curves for different stellar spin fractions, $\lambda$. Compared to a non-spinning star, prograde spins (black) lead to earlier fallback with higher peaks and retrograde spins (red) fall back later with smaller peaks. The deviation from the non-spinning case (dash-dot) increases with spin magnitude.}}
    \label{fig:fbanalytical}
\end{figure*}

The integral in Equation \eqref{mdotfb} cannot, in general, be done analytically so we numerically integrate it and plot the solutions for several $\lambda$ in Fig.~\ref{fig:fbanalytical}.

\section{Simulations}

We present simulations using the 3D smoothed particle hydrodynamics (SPH) code {\sc{phantom}} \citep{price_phantom:_2017}. We use a solar-type star modelled as a polytrope ($\gamma=5/3$) and a $10^6M_{\odot}$ black hole sink particle modelled with Newtionian gravity \footnote{Relativistic effects are small during the disruption phase for these parameters. It would be necessary to include relativistic effects to model the circularisation process, and for very deep plunges \citep{guillochon_hydrodynamical_2013,gafton_relativistic_2015} with pericentre distances $\sim1R_{\rm{g}}$.}. The first step of the simulation is to relax the polytrope to put the star in equilibrium. Once relaxed, the star is placed on a parabolic orbit starting outside the tidal field of the hole at $\sim4.9r_{\rm{t}}$ and with an impact parameter $\beta=1$, where $\beta={r_{\rm{t}}(\Omega=0)}/{r_{\rm{p}}}$. We do not use the tidal radius equation affected by spin (Equation~\ref{eq:rtspin}) as this allows us to see the effect of the stellar spin in a controlled manner, rather than changing the orbit and then trying to differentiate between the two effects. The star is initially in hydrostatic equilibrium where pressure and self-gravity are balanced. Disruption starts as the star passes within the tidal radius where the tidal force of the hole overcomes the self-gravity of the star and the pressure redistributes the energy in the star and widens the specific energy distribution \citep{lodato_stellar_2009}. The disrupted star further stretches into a stream as it continues its orbit. We can follow the return of the bound material to the hole to measure the fallback rate. The black hole is initially given a small accretion radius ($20R_{\rm{g}}$) to prevent swallowing the star whole. Once the disrupted star is sufficiently past the hole, the accretion radius is increased to $3r_{\rm{t}}$. We do not attempt to resolve the circularisation and disc-forming process so any particle that crosses the accretion radius is removed from the simulation and is considered accreted. We measure the fallback rate from the rate of particles being removed.

We set the polytrope spinning with a corotation angular velocity, equation~(\ref{eq:omgcorot}), at a fraction of its break-up velocity. The corotation corresponds to the inner and outer parts of the star rotating at the same rate but with different velocities. We look at break-up velocity fractions: $\lambda =(\pm 0.002,\pm 0.01,\pm0.05,\pm0.2)$, where positive corresponds to prograde spin and negative is retrograde. We chose these values to show that even a small spin, like the Sun ($\lambda=0.002$), can have an effect on the fallback curve. Higher spins for solar type stars do not fit with observations and some simulations of higher $\lambda$ do not get fully disrupted. Within this paper we refer to each spin by its fraction $\lambda$ and direction, and define the spin of the star as:

\begin{equation} \label{eq:omgcorot}
\Omega_{\rm{*}}=\lambda \Omega_{\rm{br}}=\lambda \sqrt{\frac{GM_{*}}{R^3_{*}}}
\end{equation}

We relax the star before starting the simulations to remove initial perturbations (we relax the star for $10$ rotation periods in the relevant corotating frame, with a velocity damping. In practise the star relaxes after $\sim 1$ rotation period).

Spin effects the stellar structure by expanding the star isotropically in the $x$ and $y$ directions but not in $z$ as there is more centrifugal support in the $z=0$ plane. This causes the star to be non-spherical in the $z-x$ plane but for low spins there is little distortion so the star is approximately spherical with an aspect ratio of $1.03$ for a spin of $\lambda=0.2$. The distortion is more significant for higher spins, e.g. for $\lambda=0.4$ (not simulated) the aspect ratio is $1.17$.

\begin{figure}
	\includegraphics[width=\columnwidth]{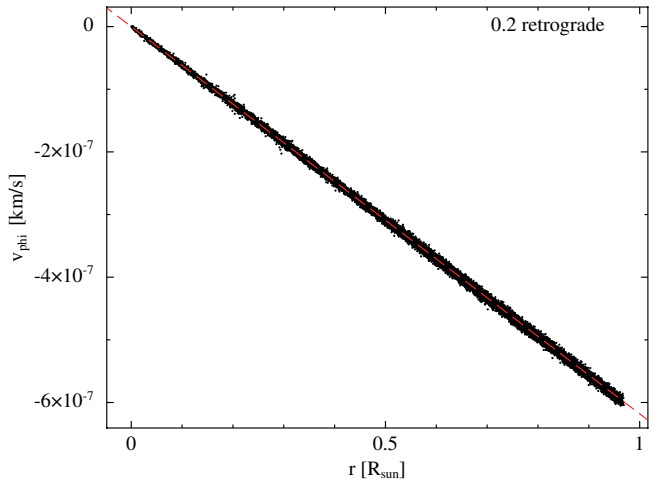}
    \caption{A check that the spin of the polytrope (black particles) maintains the correct velocity (red dashed) by taking the polytrope out of the corotating frame and putting it in isolation in its inertial frame.}
    \label{fig:vphir}
\end{figure}

We relax the spinning polytrope in the corotating frame where it appears static. We can then check the polytrope is spinning at the correct $\Omega_{\rm{*}}$ by taking it out of the corotating frame after it has relaxed, and placing it in isolation in its inertial frame. Fig.~\ref{fig:vphir} shows the azimuthal velocities of the particles (black dots) as a function of radius together with the exact solution (red dashed line), where it shows a good fit to the correct spin.

We can measure the fallback rates for each spin and compare these to the non-spinning case. We also look at the decay of the curves and whether they follow the predicted $t^{-5/3}$ behaviour. We model the evolution for $\sim 1.6$\,yrs as usually most of the bound material will fall back within this time ($\sim 76\%$ for the non-spinning case). If the power law index does not settle to the expected $-5/3$ within early times, then it is unlikely to ever be observed. The luminosity, which follows the behaviour of the fallback curve if the viscous timescale is much shorter than variations in the fallback rate \citep{lodato_challenges_2012}, will be too low to detect if $t^{-5/3}$ is not reached until late times.

The results we present in this paper were done with a million particles. We also ran the simulations with $10^4$ and $10^5$ particles and found no physical change in shape of the fallback curves, only a decrease in simulation noise and error bar size with an increase in particle number which allows us to view any real features with more clarity.

\subsection{Results}
\label{sec:simulations}

\begin{figure*}
\centering
	\includegraphics[width=0.7\textwidth]{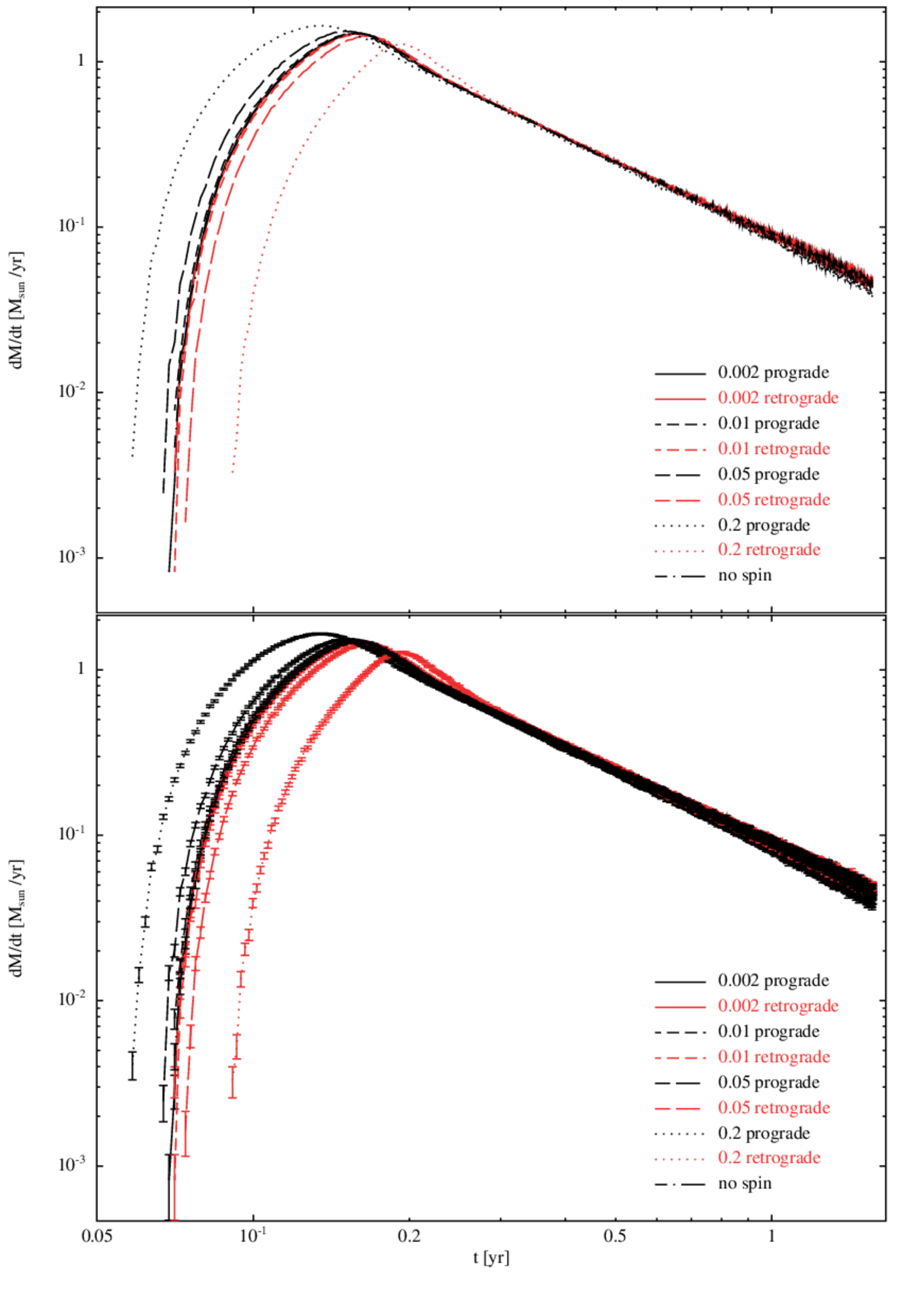}
    \caption{Comparing fallback rates of bound stellar material for different stellar rotation velocities both prograde (black) and retrograde (red) with respect to the orbit (shown with error bars in the bottom panel). We see that stars with prograde spin disrupt sooner and with higher peaks compared to the non-spinning case (dash-dot), whereas a retrograde spin hinders disruption as the tidal forces need to spin down the star first.}
    \label{fig:np1milallspins}
\end{figure*}

\begin{figure}
	\centering
	\includegraphics[width=0.5\textwidth]{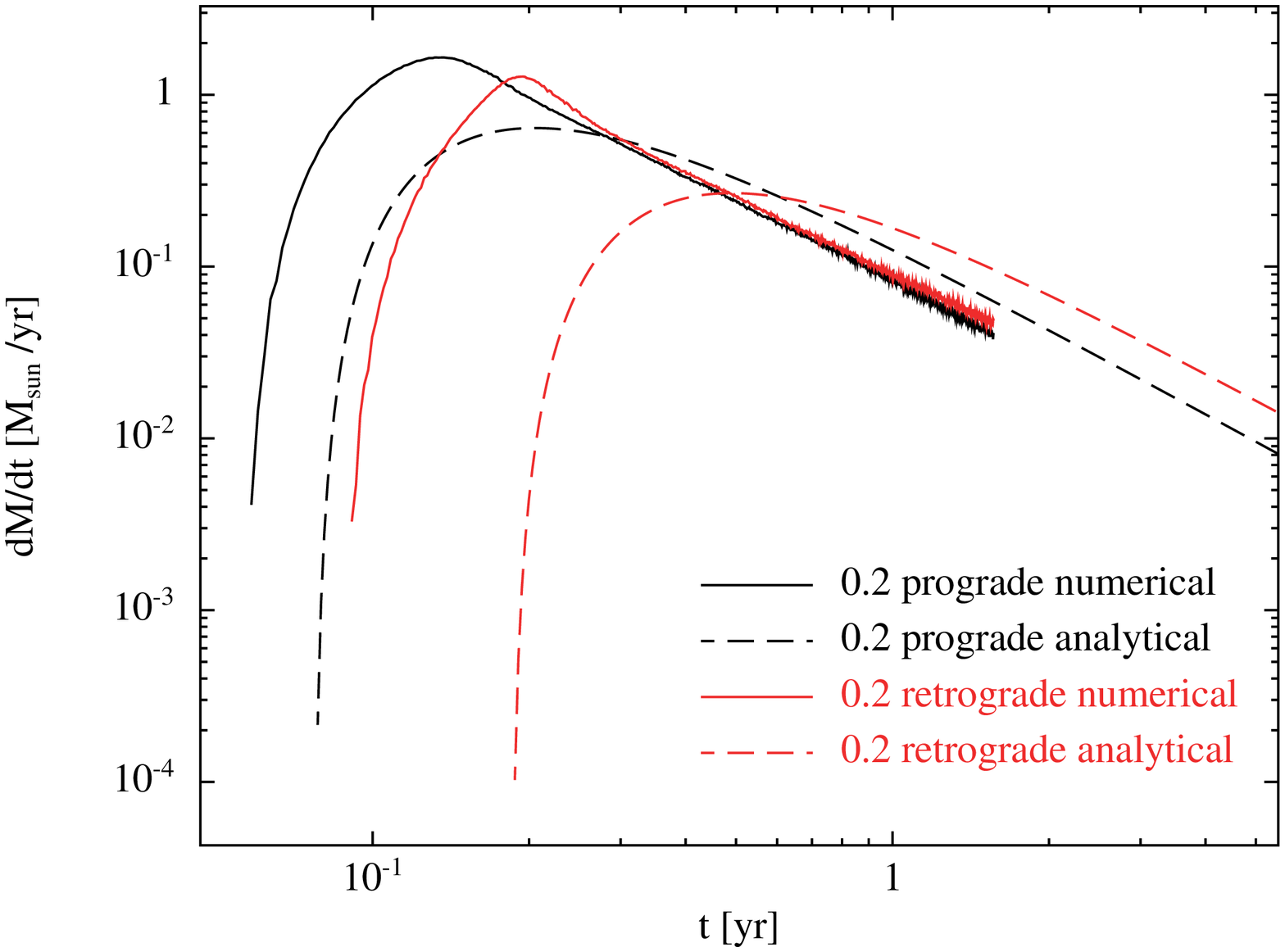}
    \caption{A comparison of the simulation fallback curves (solid lines) with analytically predicted curves (dashed) from Section 2. The analytical and numerical show the same trends of a prograde spin giving a higher peak and earlier fallback compared to retrograde. However, the differences show the need for numerical modelling of TDEs to correctly recover features in the curves from effects such ass changes to the stellar structure due to spin.}
    \label{fig:num_analytical_compare}
\end{figure}

Fig.~\ref{fig:np1milallspins} shows the fallback rates, over a time of approximately $1.6$ years, for different stellar spin velocities. The curves for the retrograde spins show less and delayed fallback of material to the hole compared to the prograde and non-spinning cases. In Fig.~\ref{fig:num_analytical_compare} we compare our numerical simulations with our analytical predictions from Section 2. Both the analytical and numerical solutions show that the fallback occurs earlier for prograde spins and later for retrograde spins. They also both show the same trend with peak fallback rate -- the prograde case has a higher peak than the retrograde case. However, the analytical and numerical curves for the same spin value do not agree closely due to the simplifying assumptions in the analytical model discussed at the start of Section 2.

\begin{figure*} 
\centering
{\includegraphics[width=0.9\textwidth, height=\textheight, keepaspectratio]{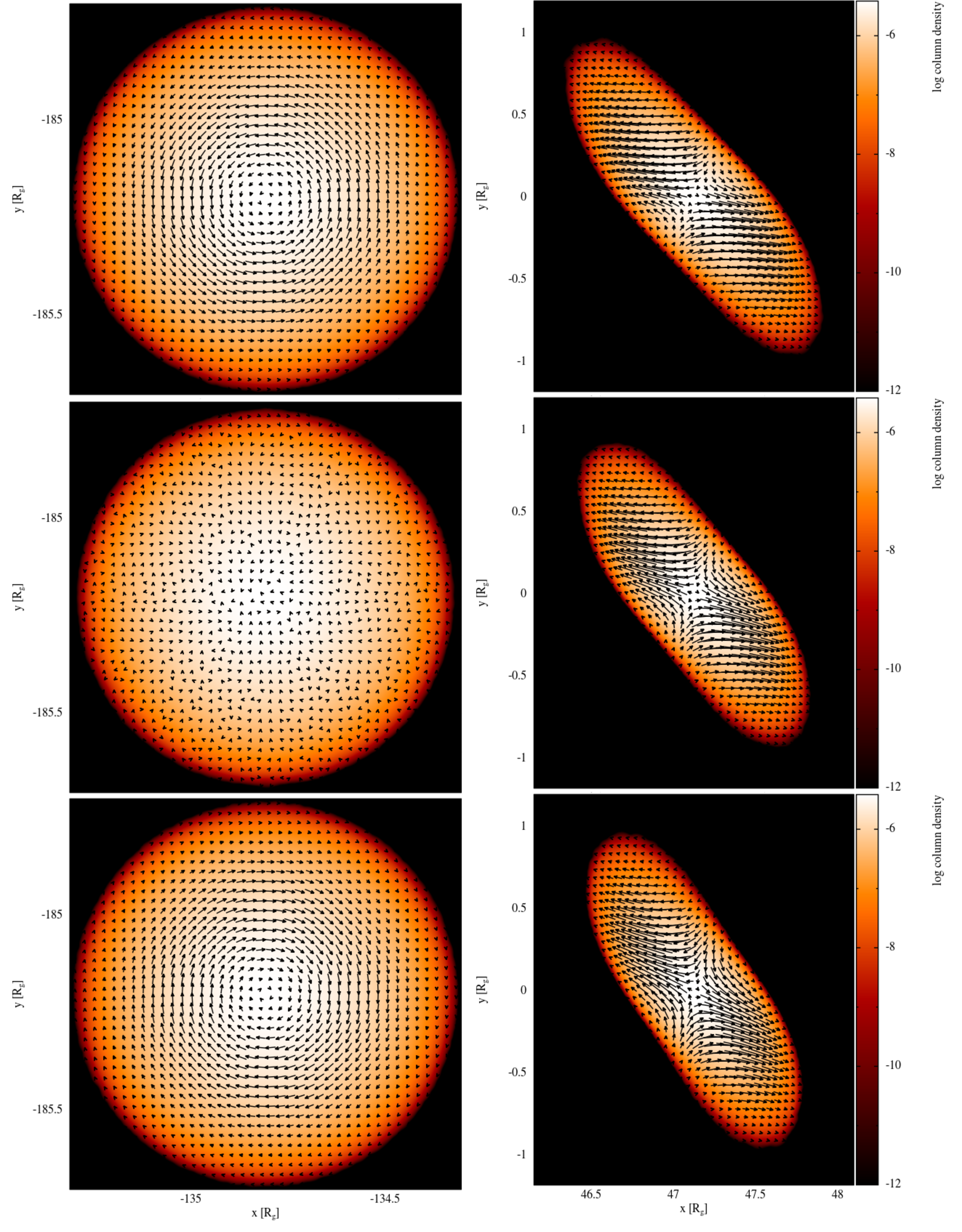}}
\caption{Star at initial position on orbit (left) and then at pericentre (right) for prograde 0.2 (top), non-spinning (middle), retrograde 0.2 cases (bottom). Overlayed is the velocity structure of the star (not including orbital velocity) and the star spins prograde or retrograde with respect to its orbit. The length of the velocity arrows indicates the column integral of $v\,{\rm d}z$}
\label{fig:TDEXY_PERI_DIS}
\end{figure*}

\begin{figure*}
\centering
{\includegraphics[width=0.8\textwidth, height=\textheight, keepaspectratio]{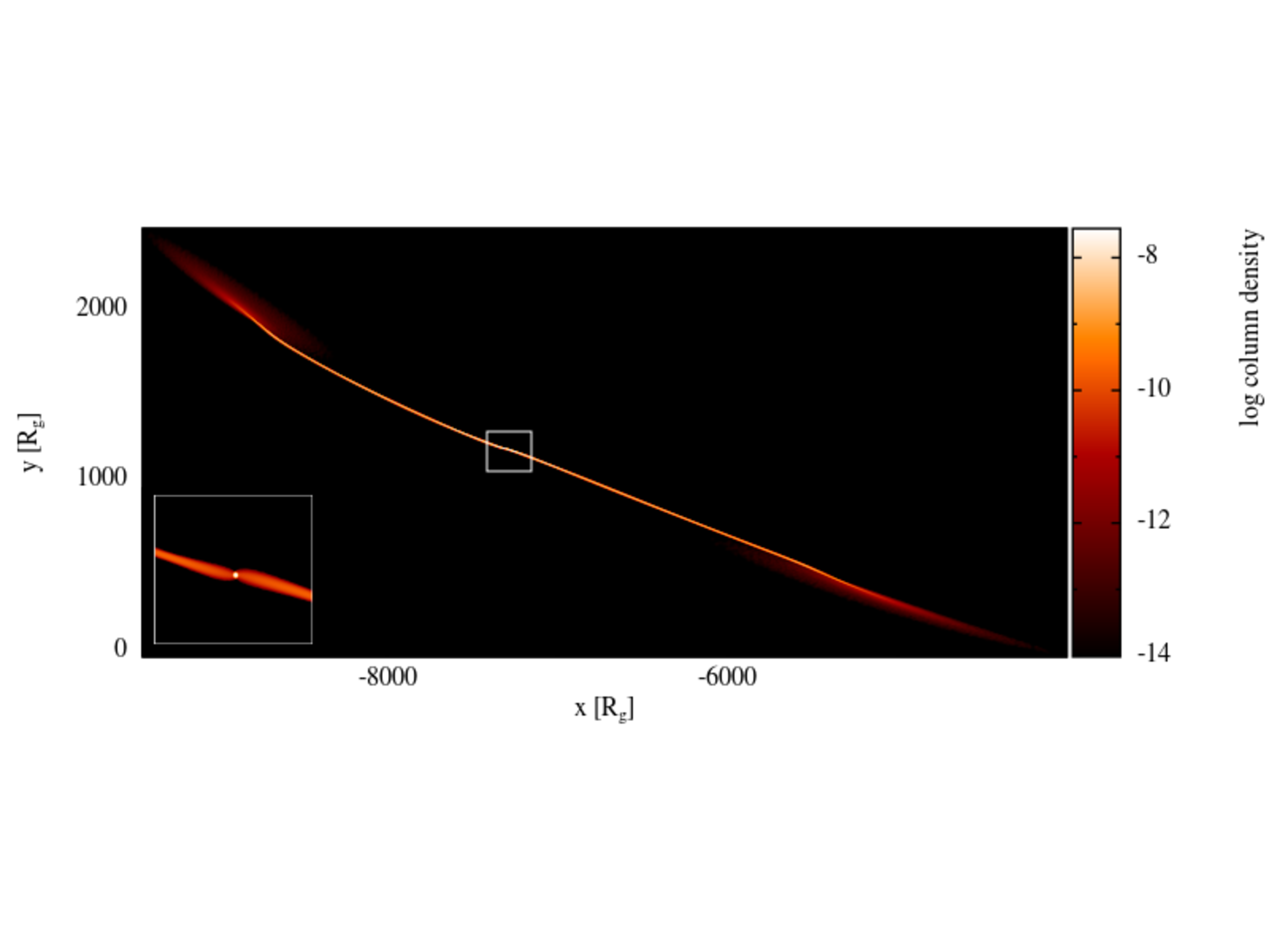}}
\caption{For the 0.2 retrograde disrupted star after passing pericentre we can see the clear effect that a retrograde spin has on the extent of disruption from pericentre and onwards, particularly where we see a small dense core still intact.}
\label{fig:TDEXY_PERI_DIS2}
\end{figure*}

\begin{figure*}
\centering
\includegraphics[width=0.8\textwidth, keepaspectratio]{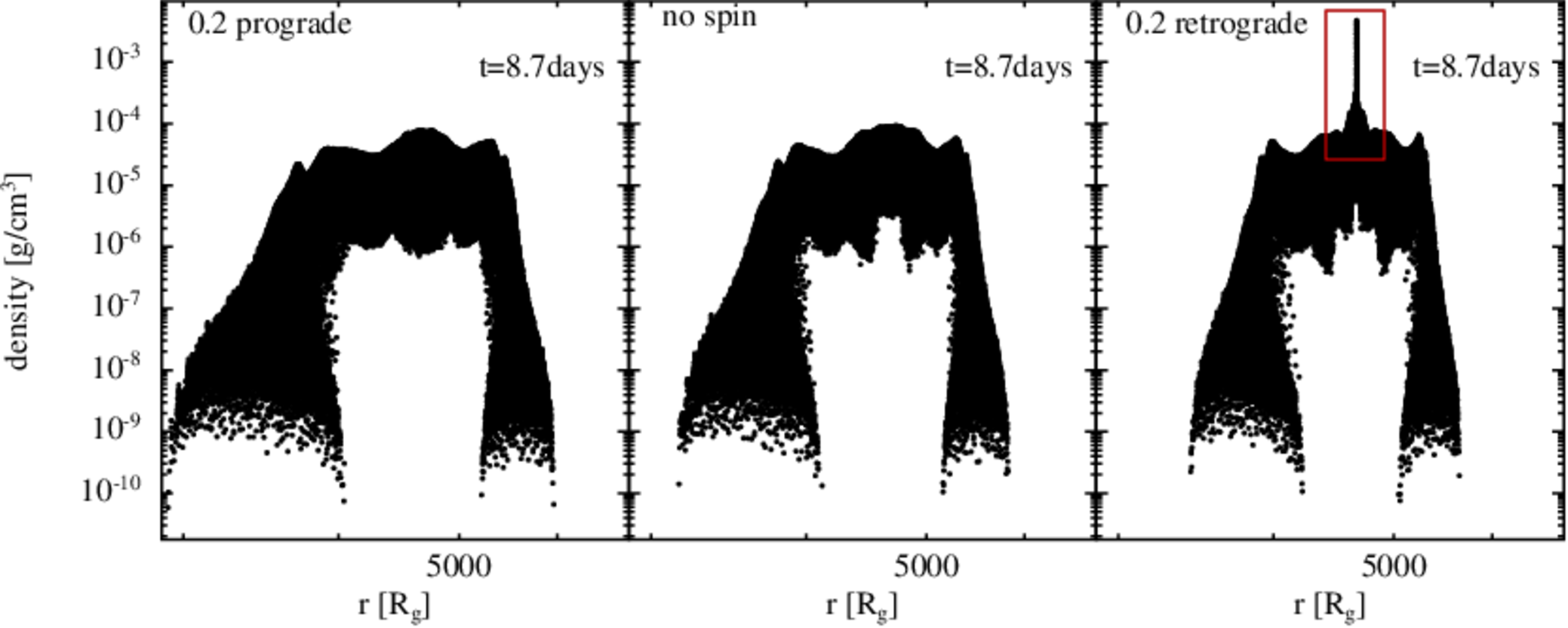}
\caption{The density structures of the disrupted stars as a function of distance from the black hole. The left and right tails of each plot show the bound and unbound parts of the debris stream. In the retrograde plot the vertical peak in the centre, highlighted within the box, shows the intact stellar core from it being harder to disrupt.}\label{fig:densitytdes}
\end{figure*}

We show in Fig.~\ref{fig:TDEXY_PERI_DIS} the velocity structure, without their orbital velocity i.e. only showing the effect of spin and the tidal force, of the star at its initial position before orbiting and at pericentre. We see that prograde is spinning in the direction of the orbit which causes it to be disrupted quicker. For faster retrograde spins (e.g. $-0.2$) the stars remained partially intact, as a result of the tidal forces having to spin down the star before disruption can occur. We can see this highlighted in the zoomed section of Fig.~\ref{fig:TDEXY_PERI_DIS2} and in the right panel of Fig.~\ref{fig:densitytdes}, where the spike in density is the bound core. This is expected as the retrograde case moves the tidal radius inwards (Equation~\ref{eq:rtspin}; recall we have used $\beta(\Omega=0)=1$ to define the simulation orbits). While a retrograde spin hinders stellar disruption, a prograde spin will quicken disruption and debris fallback as the stellar material is slightly more bound so returns to the hole quicker. 

\begin{figure*}
\centering
{\includegraphics[width=0.7\textwidth, height=\textheight, keepaspectratio]{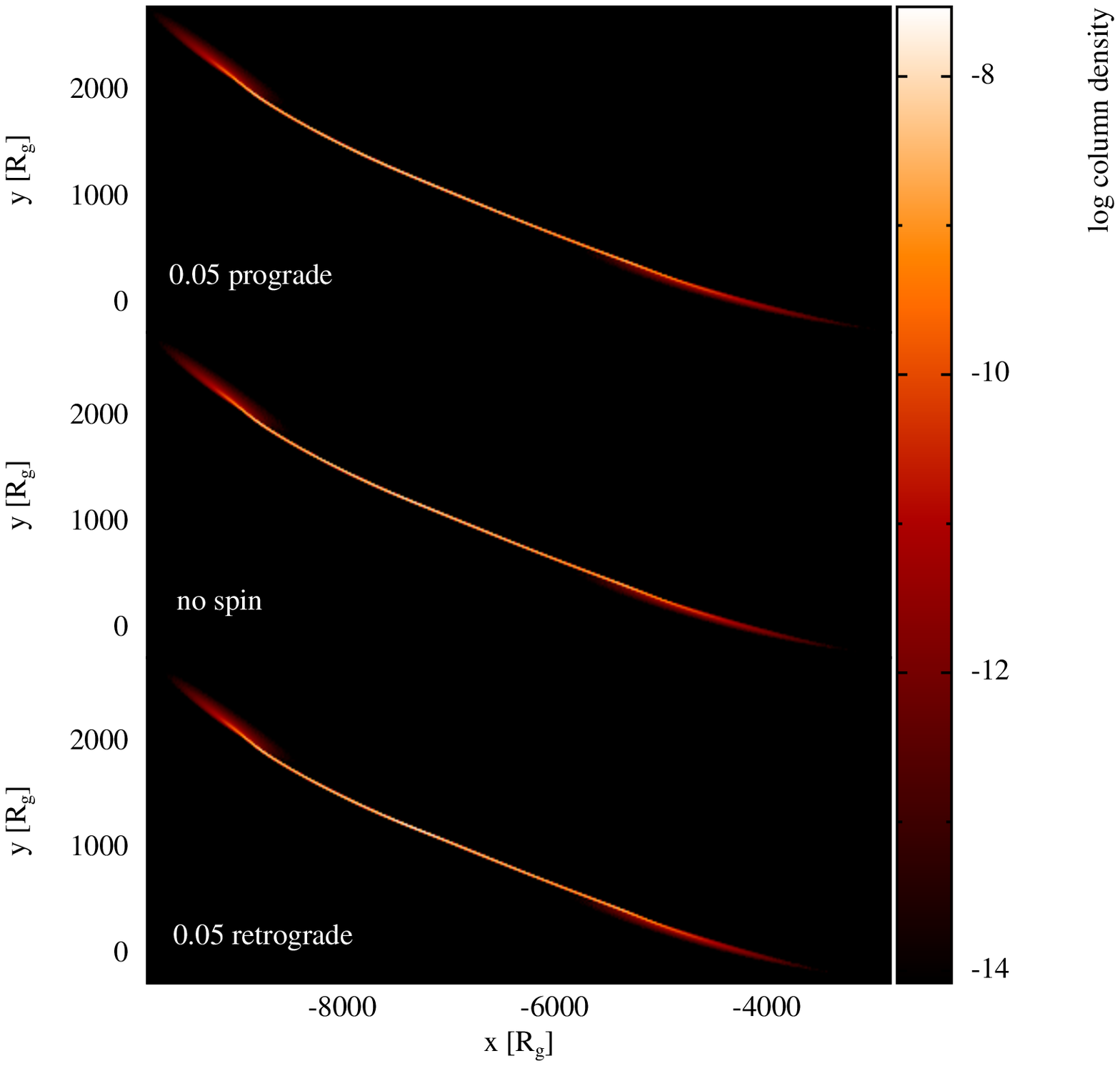}}
\caption{The disrupted debris streams for zero spin (middle panel) and $\lambda=0.05$ prograde (top) and retrograde (bottom). The stream is plotted at a time of $t = 17$ days post pericentre. Stellar spin does not significantly affect the geometry of the stream. There is a detectable difference in the streams lengths, which is due to the quickened/delayed disruption from prograde/retrograde spins.}
\label{fig:streamstructure}
\end{figure*}

\begin{figure*}
\centering
{\includegraphics[width=0.5\textwidth, height=\textheight, keepaspectratio]{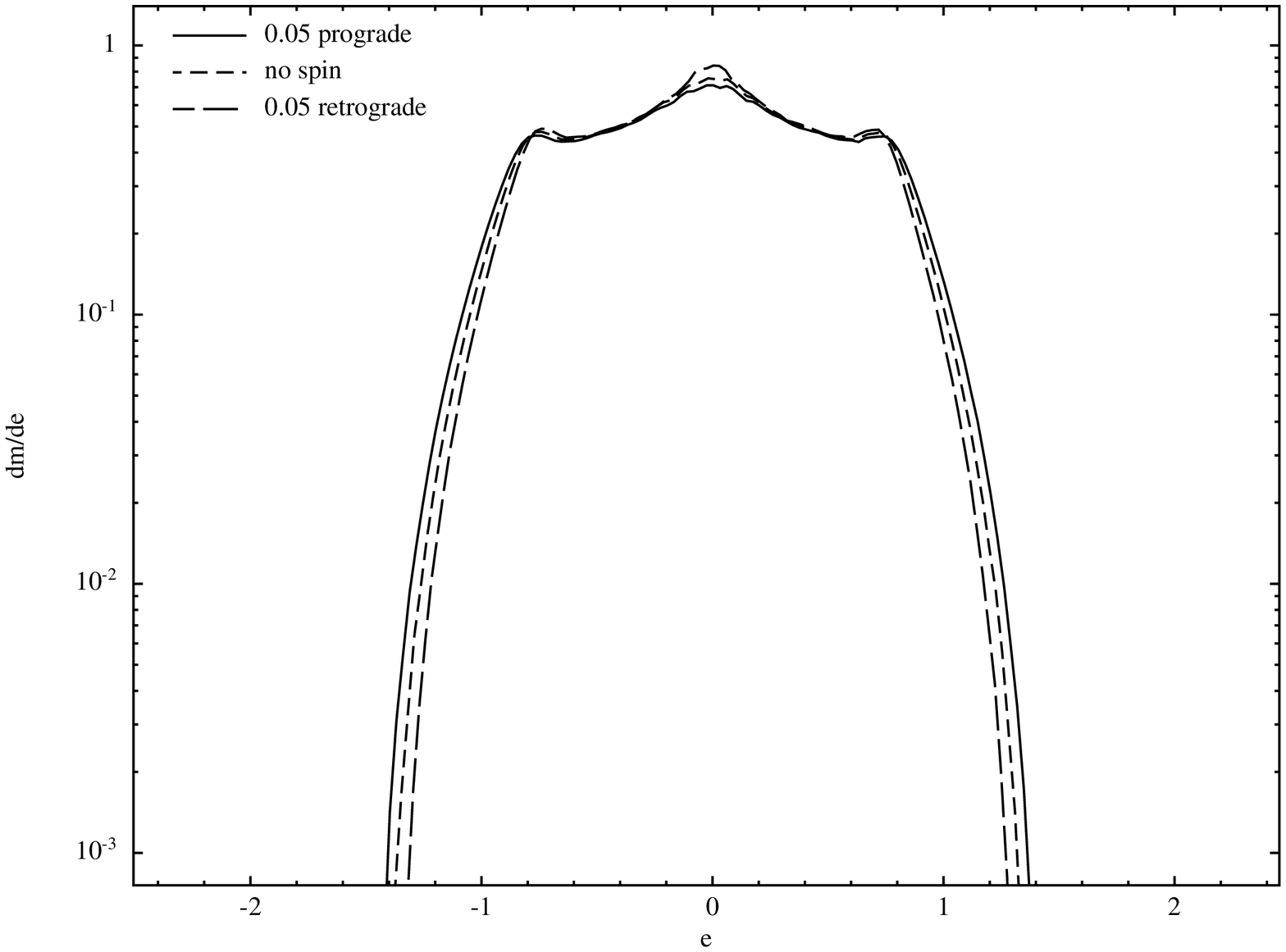}}
\caption{The energy distribution for the zero spin (short dashed) and 0.05 spin (prograde: solid and retrograde: long dashed) simulations shown in Fig~\ref{fig:streamstructure}, where $e=E/\Delta E$, $\Delta E = G M_{\rm{BH}} R_* /R^2_{\rm{peri}}$ and $dm=dM/M_*$. These  differences in the energy distribution result in differences in the fallback rates.}
\label{fig:energyspreads}
\end{figure*}

We can see from Fig.~\ref{fig:streamstructure} the stellar rotation does not have a significant impact on the stream geometry, except where a bound core survives for higher (retrograde) spins as in Fig.~\ref{fig:TDEXY_PERI_DIS2}. In Fig.~\ref{fig:energyspreads} we show how spin affects the energy distribution. As discussed in Section 2 and shown by Eq.~\ref{Eeq}, for small spins fractions the change in the energy distribution due to stellar rotation is small compared to effect from the tidal force.

\begin{figure*}
\centering
\includegraphics[width=0.9\textwidth]{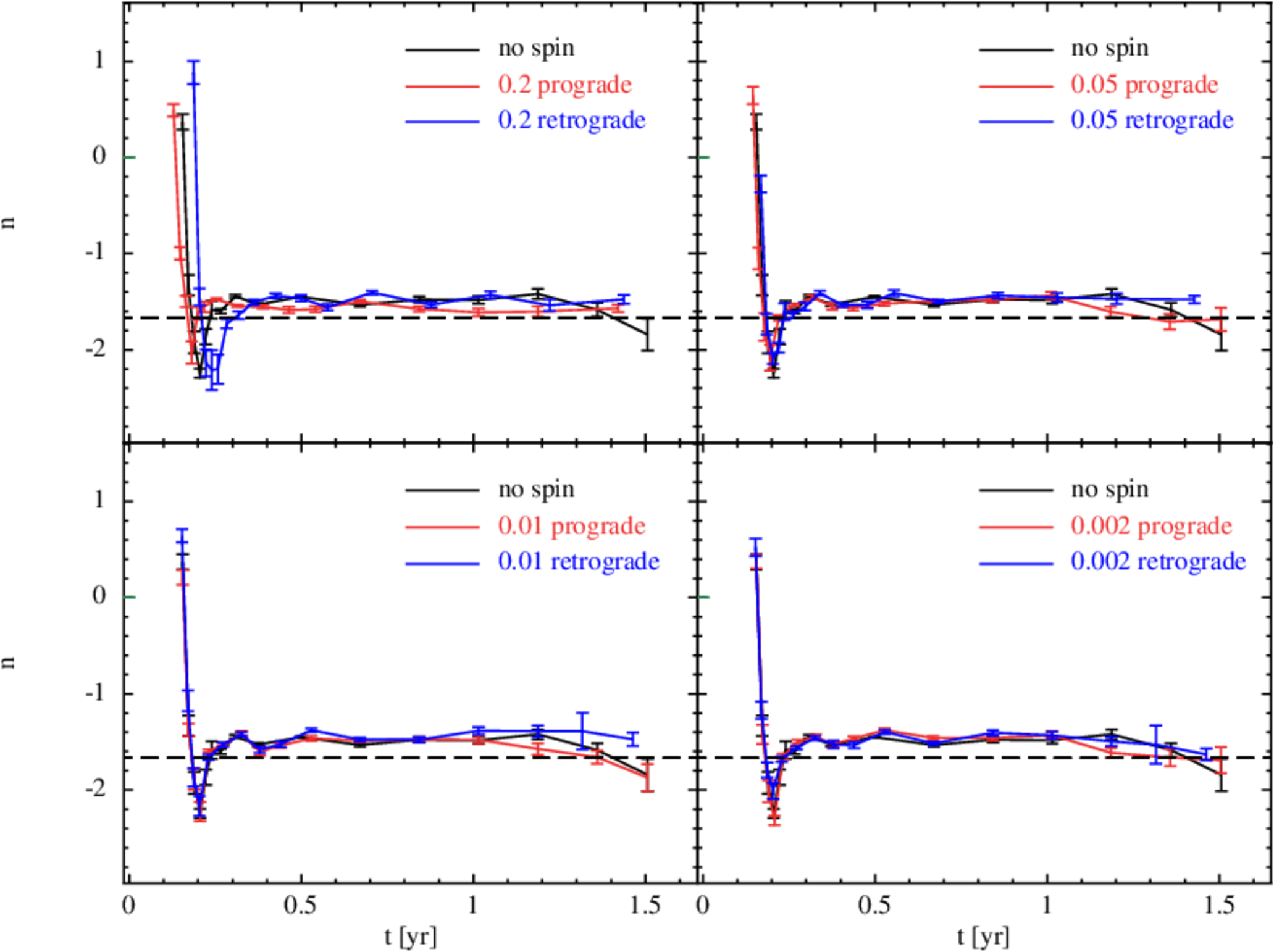}
\caption{Power law evolution for different spins plotted with the expected $-5/3$ (dashed), where $n=d \log\dot{M}/d \log t$ is the power law index. The initial overshoots of the expected $t^{-5/3}$ are due to accreting clumps which are larger for retrograde spins. The larger error bars towards the end are due to increased noise in the later parts of the fallback curves. We see that the power law does not converge on $-5/3$ for most of the TDE evolution. It is unlikely that a $t^{-5/3}$ lightcurve would be observed at late times.}
	\label{fig:pwrlaws}
\end{figure*}

We measured the power law of the fallback decay to see whether the curves in the spinning cases follow the expected $t^{-5/3}$. We performed linear regression on sections of the slope and found that the power law does vary from the predicted value, see Fig.~\ref{fig:pwrlaws}. The power law index initially overshoots the $-5/3$ and more so in the retrograde cases. This is due to the self-gravitating nature of the stream \citep{coughlin_variability_2015}. The power law may bounce back to a less steep gradient than $-5/3$ after the overshoot as material has been redistributed along the stream by self-gravity. At late times the powerlaw index is roughly constant. We can also see that the power law indexes do not settle very well on the expected $n=-5/3$, and generally stay above this until late times. At late times the errors grow as the number of particles in the stream decreases. This is in agreement with \cite{lodato_stellar_2009} who also found the power law to be shallower and only reached $t^{-5/3}$ at late times for non-spinning polytropes. This has been found in several other studies \citep{guillochon_hydrodynamical_2013,coughlin_variability_2015,2018MNRAS.478.3016W}.

We ran simulations for misaligned spins for $\lambda=0.05$ where we rotate the star to produce $\theta=45^{\circ}$ (where the angular velocity vector of the star is rotated by $45^{\circ}$ around the $x$-axis for prograde; i.e. the angular velocity vector has positive $y$ and $z$ components) and $90^{\circ}$ (where the angular velocity vector points in the direction of the orbital velocity at pericentre). We see in Fig.~\ref{fig:misalignspin} that varying the spin angle in this way only changes the fallback curve slightly compared to the aligned spins for the $45^{\circ}$ case. For the $90^{\circ}$ case, we expect from Eq.~\ref{Eeq} that there is no first order effect from spin (and remembering that any second order effect is negligible) which is confirmed by Fig.~\ref{fig:misalignspin} where the fallback curves match the non-spinning case.

\begin{figure*}
\centering
\includegraphics[width=0.7\textwidth]{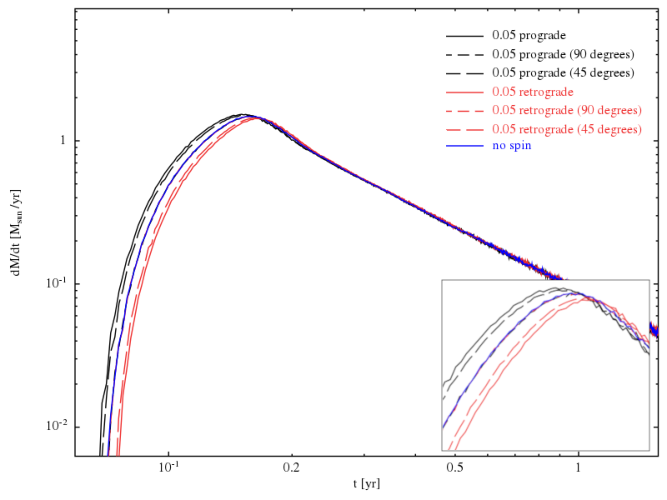}
\caption{Fallback curves for aligned (solid) and misaligned (dashed), prograde (black) and retrograde (red) spins and for the non-spinning case (blue solid). For the simulated $45^{\circ}$ angles we find only a small variation in the fallback curves compared to an aligned spin with the same magnitude. However, for both the $90^{\circ}$ cases, where the angular velocity vector is parallel to the orbital velocity vector at pericentre, the stellar spin has no effect on the stream and these fallback curves match the non-spinning case.}\label{fig:misalignedspin}
\end{figure*}\label{fig:misalignspin}

\section{Discussion and conclusions}

We ran simulations of a solar mass star, modelled as a polytrope of $1$ million particles with index $\gamma=5/3$, being tidally disrupted by a $10^6M_{\odot}$ black hole with an impact parameter of $\beta=1$ for a just full disruption. We did this for a standard case of a non-spinning star and black hole and then introduced stellar spins of values: $\pm 0.002$, $\pm 0.01$, $\pm 0.05$ and $\pm 0.2$ (fractions of the star's break up velocity, where $\Omega_*=\lambda \sqrt{GM_*/R^3_*}$) where sign indicates direction of spin with respect to its orbit, taking prograde as positive.

When we include stellar rotation into the simulation, we get interesting changes in behaviour. The direction of the star's spin will either help or hinder disruption as the tidal forces will also impart a spin onto the star. If the star is spinning prograde with respect to its orbit then it is spun up by the tidal forces and will fallback quicker and with more material for faster initial spins. Conversely, the black hole has to spin down a retrograde spinning star leading to delayed, and sometimes only partial, disruptions.

We also see some interesting features around the peak of the fallback curve, in all but the faster prograde cases, where the fallback of bound material deviates from the expected $t^{-5/3}$ decay by accreting extra mass in self-gravitating clumps in the debris stream and then less mass afterwards. The total mass accreted still averages out to half of the original stellar mass (due to half of the debris being bound, half unbound). We find that even after the overshoot occurs, the power law remains shallower and does not settle to the expected $t^{-5/3}$ until late times, by which point the TDE is unlikely to be observable anyway.

We initially calculated predictions for the debris fallback rate with the impulse approximation, which assumes the star is undisturbed until it reaches pericentre. Comparing to our numerical simulations, we see the analytical solutions are not a perfect fit with smaller peak fallback rates and longer material return times than the numerical. The impulse approximation also misses out features around the peak that should occur due to changes in the stellar structure before it reaches pericentre. However the analytical solutions recover the trends observed in the simulations, e.g. how the rise and peak fallback rates change with stellar spin.

Our analytic arguments suggested, and our numerical simulations confirmed, that stellar spin is only important for modifying the features of the fallback (e.g., the return time of the most bound debris and the time to peak fallback rate) when the star is spinning at a modest fraction of its breakup velocity. One way of generating such rapidly spinning stars prior to disruption is if the disrupting SMBH is in a binary system: as shown in \citet{coughlin_binaries_2017} a star can have a number of ``close encounters'' -- where the star comes within at least three tidal radii of either black hole -- prior to being disrupted as it traces out a chaotic, three-body orbit in the binary potential. By performing a statistical analysis of millions of three-body encounters, \citet{coughlin_binaries_2017} demonstrated that $\gtrsim 10\%$ of all three-body orbits resulting in disruptions had at least one close encounter. In these close encounters, while the tidal field of the black hole may not be sufficient to completely unbind the star, the tidal torque will spin the star up to a significant fraction ($\sim10\%$) of its breakup velocity, and repeated close encounters could push the fraction of breakup to near unity. One might therefore expect some stars disrupted by binaries to be rapidly rotating, necessitating the inclusion of this effect on the predicted fallback curves.

The change in fallback rates, both shape and normalisation, suggest that stellar spin can play an important role in defining the energy distribution of the stream and thus the observable properties of the event. It may be necessary to include such details to accurately determine system parameters from observed data.

\acknowledgments
We thank the referee for useful and constructive comments. We thank Giuseppe Lodato and Andrea Sacchi for interesting discussions on this topic.
CJN is supported by the Science and Technology Facilities Council (STFC) (grant number ST/M005917/1).  ERC acknowledges support from
NASA through the Einstein Fellowship Program, grant PF6-170150. This research used the ALICE High Performance Computing Facility at the University of Leicester. The figures were made using {\sc{splash}} \citep{price_splash:_2007}, a visualisation tool for SPH data. This work was performed using the DiRAC Data Intensive service at Leicester, operated by the University of Leicester IT Services, which forms part of the STFC DiRAC HPC Facility (www.dirac.ac.uk). The equipment was funded by BEIS capital funding via STFC capital grants ST/K000373/1 and ST/R002363/1 and STFC DiRAC Operations grant ST/R001014/1. DiRAC is part of the National e-Infrastructure.
\software{ \\{\sc{phantom}} \citep[][ http://arxiv.org/abs/1702.03930]{price_phantom:_2017}\\{\sc{Splash}} \citep[][ http://arxiv.org/abs/0709.0832]{price_splash:_2007}}

%%%%%%%%%%%%%%%%%%%%%%%%%%%%%%%%%%%%%%%%%%%%%%%%%%



\bibliographystyle{aasjournal}
\bibliography{refs.bib} 

%%%%%%%%%%%%%%%%% APPENDICES %%%%%%%%%%%%%%%%%%%%%

%%%%%%%%%%%%%%%%%%%%%%%%%%%%%%%%%%%%%%%%%%%%%%%%%%

\end{document}